\documentclass[useAMS,usenatbib]{mn2e}

\usepackage{graphicx}
\usepackage{amsmath}

\def\apj{ApJ}

\def\aa{{A\&A}}

\def\mnras{{MNRAS}}

\def\prd{{Phys. Rev. D}}
\def\physrep{{Phys.~Rep.}}   

\title[Local  dark matter clumps and  the positron excess]{Local  dark matter clumps and  the positron excess}
\author[D.T. Cumberbatch and J. Silk]{Daniel Cumberbatch$^{1}$\thanks{E-mail: dtc@astro.ox.ac.uk} and Joseph Silk$^{1}$\thanks{E-mail: silk@astro.ox.ac.uk} \\
$^{1}$Department of Astrophysics, University of Oxford, Keble Road, Oxford OX1 3RH}

\begin{document}

\date{}

\pagerange{\pageref{firstpage}--\pageref{lastpage}} \pubyear{2006}

\maketitle

\label{firstpage}

\begin{abstract}
It has been proposed that the excess in cosmic ray positrons at approximately 8\,GeV, observed on both 
flights of the HEAT balloon experiment, may be associated with the annihilation of dark matter 
within the Milky Way halo. In this paper we demonstrate how the self-annihilation of neutralino dark matter
within local substructure can account for this excess, and estimate the annihilation cross-section 
for several benchmark minimal supersymmetric (MSSM) models. We also demonstrate the effect on the permitted parameter space 
as tidal stripping effects and destruction of substructure by mergers becomes increasingly severe. 
\end{abstract}

\section{Introduction}
A large body of evidence pertaining to the existence of nonbaryonic cold dark matter (CDM) has been established over 
the past several decades, including the large-scale distribution of galaxies (\cite{abazajian}), the study of primordial light 
element abundances (\cite{cyburt04}), galactic rotation curves (\cite{begeman91}), supernova data (\cite{perlmutter99}) and the power spectrum of 
anisotropies observed in the cosmic microwave background (CMB) (\cite{bennett03},\cite{bernardis}).  At the $2\sigma$ confidence level, 
the CDM energy density parameter is now $\Omega_{\mbox{\scriptsize{CDM}}} h^2=0.113^{+0.016}_{-0.018}$ (\cite{bennett03}).  

A popular candidate for CDM is the lightest supersymmetric (SUSY) neutralino, which is a superposition of 
Higgsinos, Winos and Binos. Consequently the neutralino is electrically neutral and 
colourless, only interacting weakly and gravitationally and therefore very difficult to detect directly. 
In R-parity conserving supersymmetric models the lightest neutralino, being the lightest SUSY 
particle (LSP), is stable (\cite{weinberg}). Consequently, in a scenario where present-day CDM exists as a result of  
thermal-freeze out, the dominant species of CDM is likely to include the LSP.  The relic density of the LSP will then 
heavily depend on its mass and annihilation cross-section. The neutralino is a popular candidate for 
CDM because the expected values of these parameters are such that the corresponding relic densities are of the same order as
the currently accepted value of $\Omega_{\mbox{\scriptsize{CDM}}}$.  

In 1994 and 1995, the High-Energy Antimatter Telescope (HEAT) observed a flux in cosmic ray positrons, 
very much in excess of theoretical predictions, and peaking close to 8\,GeV (\cite{coutu99}). This observation was confirmed 
by a second flight by HEAT in 2000 (\cite{coutu01}). The effect was originally predicted in 1984 (\cite{silk84}), for both 
cosmic ray positrons and antiprotons.  While the cosmic ray antiproton feature remains elusive, at least for theorists it 
is now a well established idea that local dark matter annihilations are a possible source of the positrons contributing to 
the observed excess. Of course, improved data is urgently required to validate the significance of this observation, and 
these are expected to be taken in the near future by experiments like PAMELA, AMS-02 and Bess Polar. In this paper, 
we summarise why extant models for local dark matter annihilations fail to adequately account for the current spectral data  
on the positron feature, and present a new model.  

Studies in which the local (within several kpc) distribution of dark matter is considered to be 
smooth have concluded that the dark matter annihilation rate is insufficient to reproduce the observed positron flux. 
This naturally leads to the idea of a non-uniform, or ``clumpy'', dark matter distribution providing
the overdensities necessary to elevate dark matter annihilation rates sufficiently to produce the observed excess 
(\cite{baltz01}). The distribution of such substructure can then be estimated through the use of simulations. 
Unfortunately, while simulations generally agree on the distribution of dark matter clumps, there is a divergence 
of opinion on the survival of the lightest clumps after the effects of hierachial structure formation and tidal stripping 
are considered (\cite{diemand05}, \cite{zhao05}, \cite{diemand06}).

We note that throughout this paper we shall consider the LSP to be the lightst neutralino and therefore use the two terms 
synonymously.

\section{Positron Spectra from Neutralino Annihilation}
Positrons can be produced in several neutralino annihilation modes. For example, monoenergetic positrons can result from 
the decay of gauge bosons produced in the processes $\chi\chi\rightarrow ZZ$ or $\chi\chi\rightarrow W^+W^-$, producing 
positrons of energy $m_{\chi}/2$, where $m_{\chi}$ is the neutralino mass. A small amount of positrons can result from
the direct channel, $\chi\chi\rightarrow e^+e^-$, however such a channel seldom makes a significant impact on the overall positron flux.
A continuum of positrons, extending to much lower energies, can also be produced in the cascades of particles produced in 
annihilations. The spectrum of positrons produced in neutralino annihilations can vary significantly depending on the mass 
and annihilation modes of the LSP.  

If the neutralino is lighter than the $W^{\pm}$ and $Z$ bosons, annihilations will be dominated by the process 
$\chi\chi\rightarrow b\bar{b}$ with a minor contribution by $\chi\chi\rightarrow\tau^+\tau^-$. Assuming annihilations are 
dominated by the former process, the resulting positron spectrum will depend entirely on the mass of the LSP. For heavier 
LSPs, the annihilation products become more complex, often determined by several dominant annihilation modes including 
$\chi\chi\rightarrow W^+W^-$, $\chi\chi\rightarrow ZZ$ or $\chi\chi\rightarrow t\bar{t}$ as well as $\chi\chi\rightarrow 
b\bar{b}$ and $\chi\chi\rightarrow\tau^+\tau^-$. In our calculations, the positron spectrum per annihilation per energy 
interval $d\phi/dE_{e^+}$, uses results from PYTHIA (\cite{sjostrand01}), implemented in the DarkSUSY package 
(\cite{gondolo00}).  

\begin{figure}	
\begin{center}     \includegraphics[width=70mm,clip,keepaspectratio]{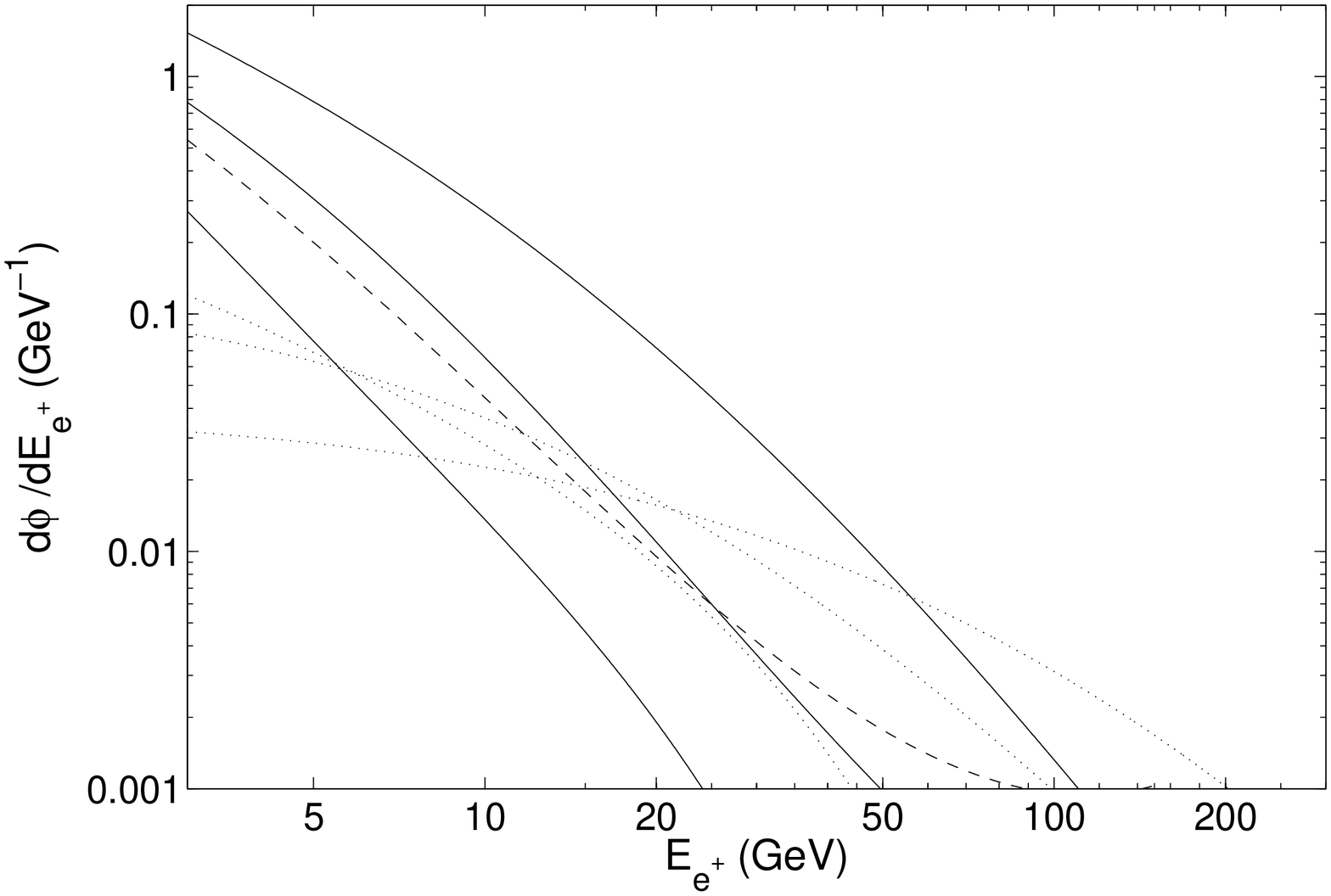}     
\end{center}   
\caption{The positron energy spectra from neutralino annihilations for modes utilised in this paper. Solid 
lines represent the positron spectra, per annihilation, for $\chi\chi\rightarrow b\bar{b}$, for neutralinos with masses of 
50 , 150 and 600\, GeV. Dotted lines are the same, but correspond to the process $\chi\chi\rightarrow
\tau^+\tau^-$.  Dashed lines correpond to the positron spectra fom the process $\chi\chi\rightarrow W^+W^-$, for a 150\,GeV 
neutralino, where the spectrum from the process $\chi\chi\rightarrow ZZ$ is very similar.}   
\label{annspectra}    
\end{figure}   

In figure\,\ref{annspectra}, we display results for positron spectra from neutralino annihilations for the most important 
annihilation modes as produced by \cite{hooper04}.  Solid lines represent the positron spectrum, per annihilation, 
for $\chi\chi\rightarrow b\bar{b}$, for LSPs with masses of 50, 150 and 600\,GeV. The dotted lines are the same, but from
the process $\chi\chi\rightarrow\tau^+\tau^-$.  Gaugino-like annihilations typically produce a spectrum which is dominated 
by $b\bar{b}$ at low energies, with contributions from $\tau^+\tau^-$ only becoming important at energies above about half 
the LSP mass. For neutralinos with a dominant higgsino component, annihilations to gauge bosons often dominate. Dashed lines
represent positron spectra resulting from the process $\chi\chi\rightarrow W^+W^-$ for a 150\,GeV neutralino, where the 
spectrum corresponding to the process $\chi\chi\rightarrow ZZ$ is very similar.  
 
\section{Positron Propagation Model}
Cosmic ray positrons being electrically charged particles diffuse through the interstellar medium (ISM) under the 
electromagnetic influence of the galactic magnetic field and background radiation. The spatial distribution of the 
galactic magnetic field is extremely complex, resulting in positron trajectories which are well approximated to 
random walks. Whilst traversing the ISM, the positrons radiate through synchrotron losses owing, to the surrounding 
magnetic fields, and by inverse Compton scattering off of starlight and CMB background radiation (\cite{webber92}).  

The diffusion-loss equation describing the above process is given by  

\begin{eqnarray} \frac{\partial}{\partial t} \frac{\partial n_{e^+}}{\partial E_{e^+}} & = &  
\nabla \cdot \left[K(E_{e^+},{\bf r})\nabla\frac{\partial n_{e^+}}{\partial E_{e^+}}\right] \nonumber\\
& & +\frac{\partial}{\partial E_{e^+}} \left[b(E_{e^+},{\bf r})\frac{\partial n_{e^+}}{\partial E_{e^+}}\right] + 
Q(E_{e^+},{\bf r}), 
\label{1} 
\end{eqnarray}   

\noindent where $\partial n_{e^+}/\partial E_{e+}$ is the number density of positrons per unit energy interval, 
$K(E_{e^+},{\bf r})$ is a diffusion coefficient, $b(E_{e^+},{\bf r})$ is the rate of energy loss  
and $Q(E_{e^+},{\bf r})$ is the source term.  

We parameterize the 
diffusion coefficient and the rate of energy loss as follows  

\begin{equation} 
K(E_{e^+})=K_0(3^{\alpha}+E_{e^+}^{\alpha})\approx 3 \times 10^{27}(3^{0.6}+E_{e^+}^{0.6}) \;\, \mbox{cm}^2\mbox{s}^{-1} 
\label{2} 
\end{equation}  

\noindent and 

\begin{equation} 
b(E_{e^+})=\tau_E E_{e^+}^2\approx 10^{-16}E_{e^+}^2 \;\, \mbox{s}^{-1}, 
\label{3} 
\end{equation}  

\noindent based upon measurements of stable nuclei in cosmic rays (primarily by fitting to observations of the 
boron to carbon ratio) (\cite{maurin02}), where in equations (\ref{2}) and (\ref{3}) we have assumed a spatially uniform 
galactic magnetic field. We then define $K$ and $b$ to be constant within a ``diffusion zone'', which here is considered to be 
a radially-infinite cylindrical slab of thickness $2L\approx$4\,kpc, which is the best-fit to observations (\cite{maurin02}, 
\cite{webber92}).

We solve equation (\ref{1}) for the local positron flux, using the proceduce described in the appendix to this paper. 

\section{Positron and Electron Backgrounds}
In order to directly compare our results to the observations made by HEAT, we must calculate the ratio of 
the local positron flux to the combined local positron and electron fluxes, called the ``positron fraction''. In order 
to do this we require
the background spectra for secondary positrons, primary electrons and secondary electrons, where we propose that the primary 
positron component originates from the neutralino annihilations. In the following, we 
make use of the parameterised fits to these spectra calculated by \cite{edsjo98} and stated here (in units 
of $\mbox{GeV}^{-1}\,\mbox{cm}^{-2}\,\mbox{s}^{-1}\,\mbox{sr}^{-1}$), as follows:  

\begin{eqnarray} 
\left( \frac{d\Phi}{dE_{e^+}} \right)_{\mbox{\scriptsize{prim. $e^-$}}} & = 
& \frac{0.16\epsilon^{-1.1}}{1+11\epsilon^{0.9}+3.2\epsilon^{2.15}}\\ 
\left( \frac{d\Phi}{dE_{e^+}} \right)_{\mbox{\scriptsize{sec. $e^-$}}} & = 
& \frac{0.70\epsilon^{0.7}}{1+110\epsilon^{1.5}+600\epsilon^{2.9}+580\epsilon^{4.2}}\\ 
\left( \frac{d\Phi}{dE_{e+}} \right)_{\mbox{\scriptsize{sec. $e^+$}}} & = 
& \frac{4.5\epsilon^{0.7}}{1+650\epsilon^{2.3}+1500\epsilon^{4.2}} 
\label{22} 
\end{eqnarray}  

\noindent again where $\epsilon =E_{e^+}/\left(1\,\mbox{GeV}\right)$. These equations agree with their respetive 
observational results to within 10-15\% for the relevent energy intervals. (For a more detailed account of the precision 
of these equations, refer to \cite{strong01}.)  The positron fraction is then calculated using

\begin{equation} 
\frac{\left( \frac{d\Phi}{dE} \right)_{\mbox{\scriptsize{prim.}}\, e^+} + \left( \frac{d\Phi}{dE} \right)_{\mbox{\scriptsize{sec.}}\, e^+} }
{\left( \frac{d\Phi}{dE} \right)_{\mbox{\scriptsize{prim.}}\, e^-} +\left( \frac{d\Phi}{dE} \right)_{\mbox{\scriptsize{sec.}}\, e^-} +\left( \frac{d\Phi}{dE} \right)_{\mbox{\scriptsize{prim.}}\, e^+} +\left( \frac{d\Phi}{dE} \right)_{\mbox{\scriptsize{sec.}}\, e^+} }. 
\label{23} 
\end{equation}  

\section{Dark matter Substructure}
The standard cosmological model assumes that all structure in the universe originated from small amplitude quantum 
fluctuations during an epoch of inflationary expansion shortly after the big bang. The 
linear growth of the resulting density fluctuations is then completely determined by their initial power spectrum 

\begin{equation}
P(k)=k^n,
\label{7a}
\end{equation}

\noindent where for $n\ge1$, clumps are formed with a wide range of scales $\propto k^{-1}$. During the expansion of the universe, smaller 
clumps coalesce to form larger ones in a process of ``bottom-up'' hierarchical structure formation, in which the mass
distribution at any given redshift can be determined through the use of numerical simulations.

In this paper we utilise the mass distribution of clumps deduced from the high-resolution simulations conducted by Diemand et al. (2005), 
of the form $dn_D/d$log$(M/M_{\odot})\propto (M/M_{\odot})^{-1}{\rm exp}[-(M/M_{\rm cut-off})^{-2/3}]$. Diemand et al. deduced
a lower mass cut-off $M_{\rm cut-off}\approx 8.03\times10^{-6}\,M_{\odot}$, however we shall also investigate the effect of using a much 
larger value, of order $10^6\,M_{\odot}$ (similar to the typical mass resolution of conventional large-scale numerical 
simulations), to simulate the destruction of lighter clumps by merging. We normalise the 
distribution so that the number density of clumps between $10^{-6}M_{\odot}$ and $10^{-5}M_{\odot}$ is 
500\,pc$^{-3}$, with a halo-to-halo scatter factor of approximately 4, as stated by Diemand et al.. 
We also invoke an upper mass cut-off, $M_{\rm max}\sim 10^{10}M_{\odot}$, defined so that heavier clumps possess 
a number density of less than 1 per unit volume within the region in which positrons contribute to the 
local flux.

The rate of dark matter annihilations within a clump crucially depends on its density profile. In this study
we adopt the widely used Navarro, Frenk \& White (hereafter, NFW) density profile (\cite{nfw})

\begin{equation}
\rho(r)=\frac{\rho_0}{c(r/r_{200})[1+c(r/r_{200})]^2},
\label{8}
\end{equation}

\noindent which is consistent with simulated results for the outer regions large-scale halos. 
The NFW profile also provides a good fit to those of the lightest clumps produced in the simulations by Diemand et al. 
(\cite{diemand05}). The maximum radius of a clump is defined here to be the radius, $r_{200}$, at which its density is 
equal to 200 times the critical density. The
concentration parameter $c$ is a further degree of freedom, found to lie between $1.6 < c < 3$ for the clumps in 
Diemand et al.'s simulations. Since the relationship beteween clump mass and concentration is complex, we approximate 
that all clumps have the same concentration and we compare results when using $c=1.6\,{\rm and}\,3$.

Since the simulations conducted by Diemand et al. were terminated at $z=26$, even though there is reason to believe that
the shape of the mass distribution of clumps is similar today for large mass scales (\cite{diemand05}), we must also 
consider mass losses from tidal stripping 
by stars as clumps traverse the galactic halo. In this study we consider a scenario where the fractional mass loss 
experienced by each clump, over a given duration, is a mass-independent constant $1-f$ and we compare results when using 
several different values of this parameter. 

The notion of a universal tidal stripping parameter is consistent with the calculation by Zhao et al., who demonstrates that 

\begin{equation}
{\rm ln}\,\frac{1}{f}\propto \frac{\langle\rho_{\ast}\rangle}{\rho_{\rm core}^{1/2}}\Delta t,
\label{9}
\end{equation}

\noindent where $\langle\rho_{\ast}\rangle$ is the average stellar densiity along the clump orbit, 
$\rho_{\rm core}$ is an effective ``core'' density of the clump and $\Delta t$ is the time period considered.
Since the majority of clumps have similar orbital parameters (\cite{zhao05}), and since we have adopted a 
universal clump profile, equation (\ref{9}) implies that $f$ is mass-independent.

\section{Positron Spectra From Dark Matter Substructure}
We consider four different neutralino models: 

First, a 50\,GeV neutralino with an annihilation branching ratio of 0.96 to $b\bar{b}$ and 0.04 to $\tau^+\tau^-$. Such a 
particle could be gaugino-like or higgsino-like, since for masses below the gauge boson masses, these modes dominate for 
either case (designated as model\,1).  

Second, we consider two cases for a 150\,GeV neutralino. One which annihilates as described in model\,1 
(designated as model\,2), and 
another which annihillates entirely to gauge bosons, $W^+W^-$ or $ZZ$ (designated as model\,4). Such neutralinos are 
typically gaugino-like and 
higgsino-like respectively.  

Finally we consider heavy, 600\,GeV neutralinos, which annihilate to $b\bar{b}$ with a ratio 
of 0.87 and to $\tau^+\tau^-$ or $t^+t^-$ the remaining time (designated as model\,3). 
(We note that even though we do not explicitely calculate results for a model involving 600\,GeV Higgsino-dominated 
neutralinos, which primarily decay to gauge bosons and partially to Higgs bosons, the resulting positron spectra per 
annihilation is likely to be very similar to that of model\,3).

Although these models do not fully encompass the extensive 
parameter space available to neutralinos at present, they do describe effective MSSM benchmarks. Furthermore, the relevant 
results for neutralinos with a mixture of the properties of those above can be inferred by interpolating 
between those presented.  

\begin{figure}	
\begin{center}     
\includegraphics[width=60mm,height=42mm,clip]{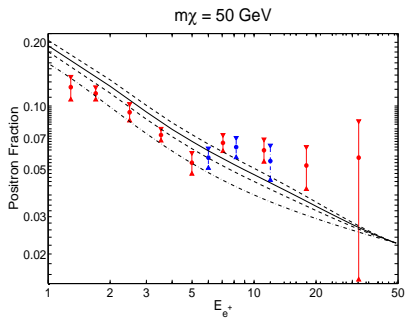}     
\end{center}   
\caption{The calculated positron fraction as a function of positron energy (in GeV), for a 50\,GeV 	 
neutralino which annihilates to $b\bar{b}$ 96\% of the time and $4\%$ to $\tau^+\tau^-$ (designated as model\,1). 
The error bars displayed are for the 1994-95 (red) and 2000 (blue) HEAT data. The solid line represents the spectra 
which best fits the data. The best-fit $\chi^2$ for this model is approximately 5.4, (for 12 data points). Dashed 
lines represent the spectra corresponding to the 1\,$\sigma$ fits to the observations, where the normalisation of the 
positron flux was considered to be a free paramter. The dot-dashed line represents the non-exotic background contribution 
to the positron fraction.}   
\label{figure2}    
\end{figure}   

\begin{figure}	
\begin{center}     
\includegraphics[width=60mm,height=42mm,clip]{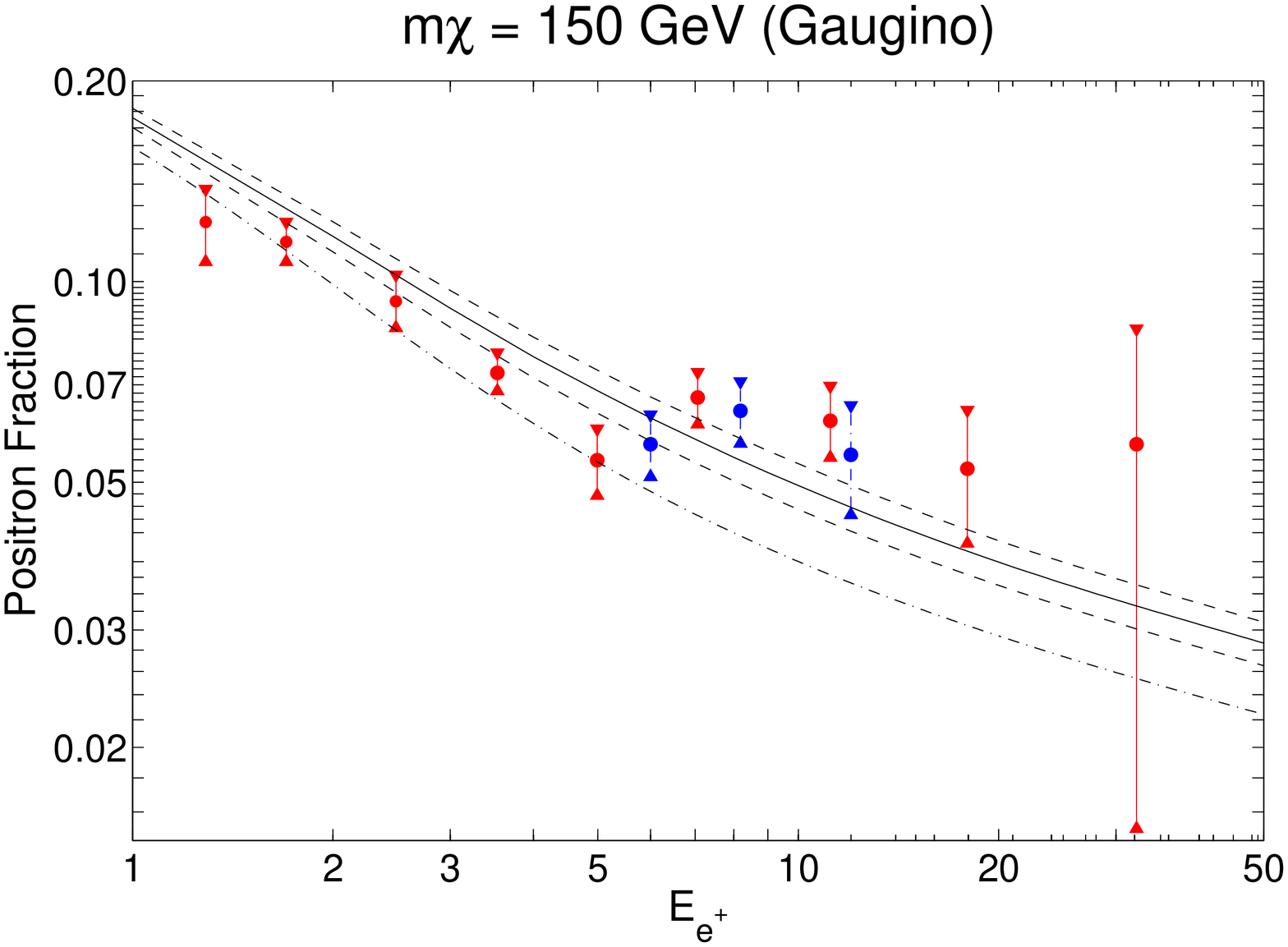}     
\end{center}   
\caption{The calculated positron fraction as a function of positron energy (in GeV), for a 150\,GeV neutralino  
which annihilates to $b\bar{b}$ 96\% of the time and $4\%$ to $\tau^+\tau^-$ (designated as model\,2). The best-fit 
$\chi^2$ for this model is 4.4, (for 12 data points). Otherwise, the same as in figure 2.}   
\label{figure3}    
\end{figure}   

\begin{figure}	
\begin{center}     
\includegraphics[width=60mm,height=42mm,clip]{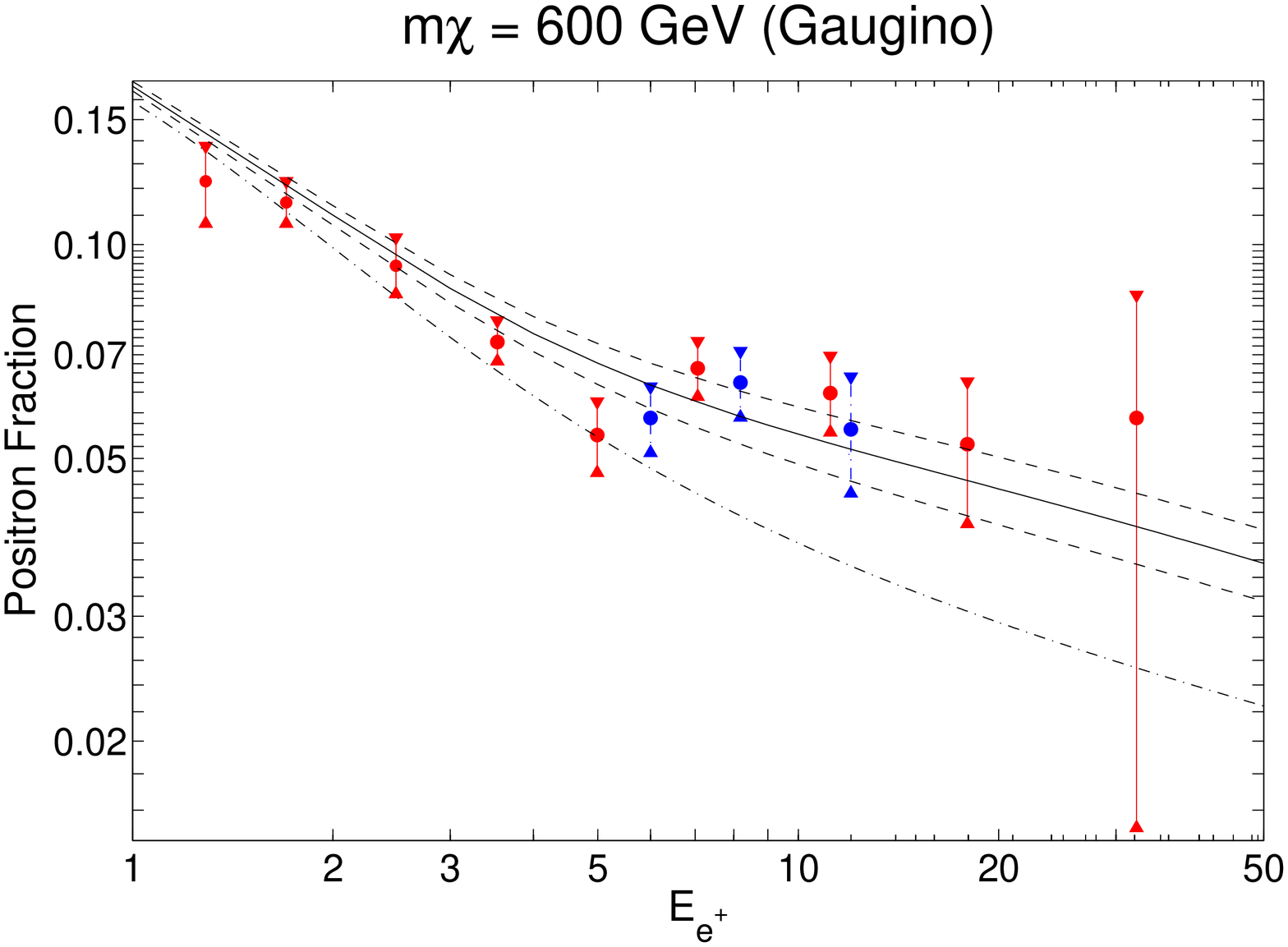}     
\end{center}   
\caption{The calculated positron fraction as a function of positron energy (in GeV), for a 600\,GeV neutralino which 
annihilates 87\% to $b\bar{b}$ and 13\% to 	 $\tau^+\tau^-$ or $t^+t^-$ (designated as model\,3). The best-fit 
$\chi^2$ for this model is 2.9, (for 12 data points).  Otherwise, the same as in figure 2.}   
\label{figure4}    
\end{figure}   

\begin{figure}	
\begin{center}     
\includegraphics[width=62mm,height=45mm,clip]{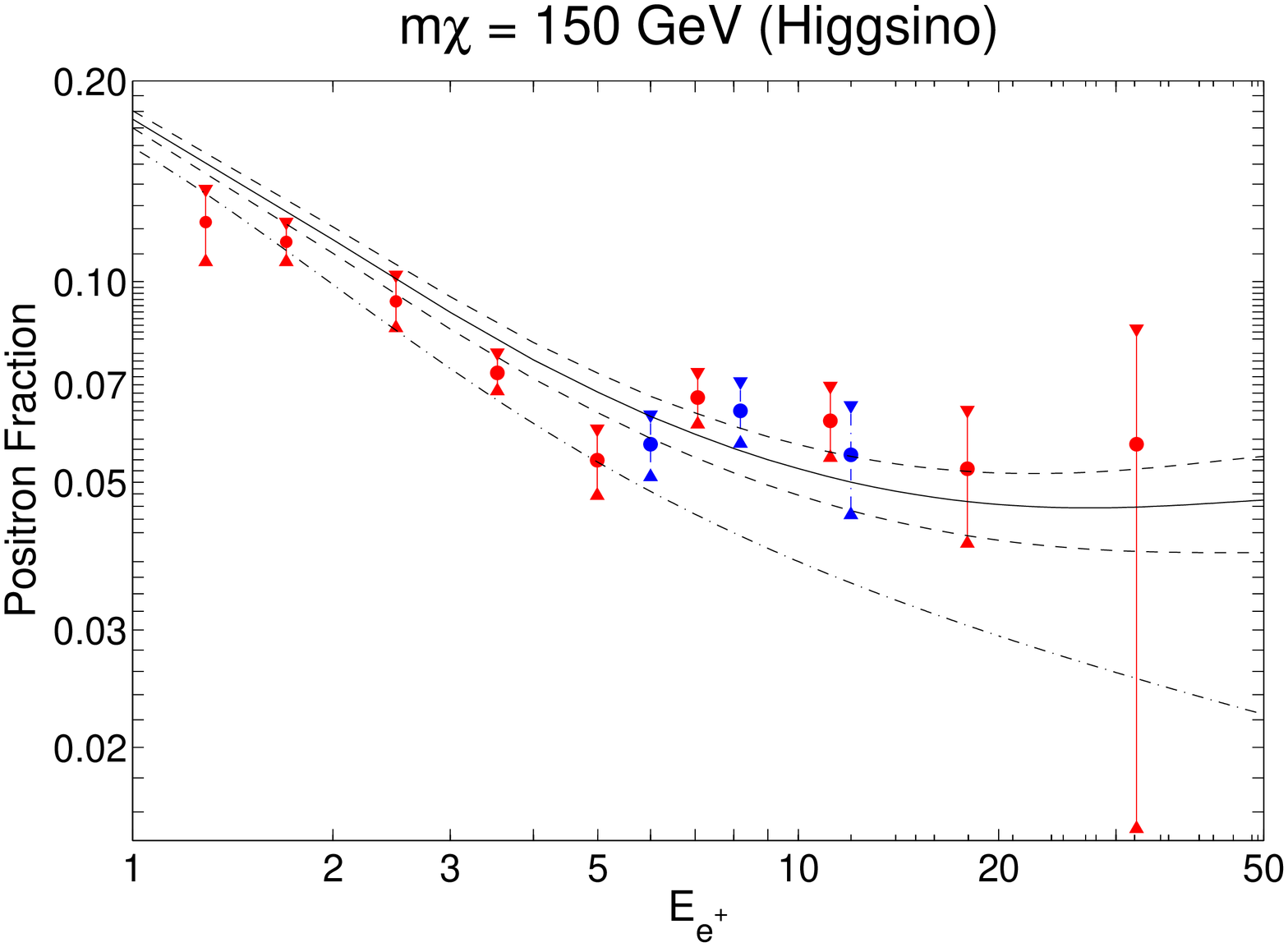}     
\end{center}   
\caption{The calculated positron fraction as a function of positron energy (in GeV), for a 150\,GeV neutralino which  
annihilates entirely to $W^+W^-$ or $Z^+Z^-$ (designated as model\,4). The best-fit $\chi^2$ for this model is 3.3,  
(for 12 data points). Otherwise, the same as in figure 2.}   
\label{figure5}    
\end{figure}   

Figures\,\ref{figure2}-\ref{figure5} display the positron fraction as a function of positron energy, calculated for each of
the four MSSM models outlined above.

The thermally-averaged product of the low-velocity annihilation cross-section and relative speed of dark matter 
particles,  
$\langle\sigma\upsilon\rangle$, was left as a free parameter and varied to fit our results to observations. 
In each figure, the solid line displays the positron fraction which fitted best to $1\,\sigma$ error bars of the 94-95 (red)
and 2000 (blue) HEAT data. Dashed lines correspond to positron fractions where $\chi^2$ (for the 12 data points), 
differs by unity from the corresponding best-fit value (1$\sigma$ results). Finally, the dot-dashed lines corresponds to 
the non-exotic background spectrum. The results are summarized in tables \ref{table1} and \ref{table2}.

\begin{table}  
\begin{center} 
\begin{tabular}{ccccc} 
\hline\hline & c=3 &  & c=1.6 &  \\
 f & $\frac{\mbox{Best-Fit}}{10^{-29}}$ & $\frac{1\sigma}{10^{-29}}$ &$\frac{\mbox{Best-Fit}}{10^{-29}}$ & $\frac{1\sigma}{10^{-29}}$\\ 
\hline\hline\\
Model 1 &&&&\\ 
1.0& 1.40& 0.55-2.05 & 3.99 & 2.51-5.48\\ 
0.1& 2.30& 1.39-3.21 & 6.61 & 4.19-9.03\\ 
0.01&5.38& 3.41-7.36 & 16.61 & 10.54-22.68 \\
\\
[0.5ex]

Model 2 &&&&\\ 
1.0& 2.99& 1.92-4.07& 7.46 & 4.84-10.09\\ 
0.1& 4.42& 2.88-5.95& 12.37& 8.10-16.65\\
0.01& 10.08& 6.58-13.58& 31.09 &20.34-41.84\\
\\
[0.5ex]

Model 3 &&&& \\
1.0& 10.90& 6.47-15.34& 31.47 & 21.36-41.58\\ 
0.1& 18.04& 11.82-24.26& 52.05& 35.60-68.50\\ 
0.01& 42.42& 28.99-55.85& 131.90& 90.21-173.59\\
\\
[0.5ex]

Model 4 &&&&\\
1.0& 3.81& 2.56-5.06& 8.73 & 5.44-12.02\\ 
0.1& 5.56& 3.76-7.36& 15.68& 10.44-20.92\\
0.01& 12.52&8.20-16.85&39.82&26.77-52.86\\
\\
[0.5ex]
\end{tabular} 
\caption{Values of the product $\langle\sigma\upsilon\rangle$, in cm$^3$\,s$^{-1}$, which give rise to 
positron fractions which best-fit the HEAT observations, for the four MSSM models considered, for 
$M_{\rm cut-off}=8.03\times10^{-6}$ and various values of $c$ and $f$. One sigma error ranges are also 
displayed.} 
\label{table1} 
\end{center} 
\end{table}   

\begin{table}  
\begin{center} 
\begin{tabular}{ccccc} 
\hline\hline & c=3 &  & c=1.6 &  \\
 f & $\frac{\mbox{Best-Fit}}{10^{-29}}$ & $\frac{1\sigma}{10^{-29}}$ &$\frac{\mbox{Best-Fit}}{10^{-29}}$ & $\frac{1\sigma}{10^{-29}}$\\ 
\hline\hline\\
Model 1 &&&&\\ 
1.0& 6.52& 4.07-8.98 & 16.14 & 10.05-22.24\\ 
0.1& 9.57& 6.05-13.08& 26.68 & 16.39-36.97\\ 
0.01&21.90& 13.81-29.99& 67.69 & 42.95-92.44\\
\\
[0.5ex]

Model 2 &&&&\\ 
1.0& 11.75& 7.63-15.87 & 28.87& 18.48-39.26\\ 
0.1& 17.13& 11.22-23.04& 48.47& 31.67-65.28\\
0.01& 36.97& 22.43-51.52& 121.96&79.81-164.10\\
\\
[0.5ex]

Model 3 &&&& \\
1.0& 47.73& 31.16-64.30& 122.08& 83.55-160.60\\ 
0.1& 72.36& 48.95-95.74& 201.55& 137.95-265.14\\ 
0.01& 163.78& 111.20-215.35& 508.81& 348.73-668.90\\
\\
[0.5ex]

Model 4 &&&&\\
1.0& 12.27& 6.94-17.61& 36.76& 24.69-48.83\\ 
0.1& 20.85& 13.39-28.31& 60.95& 41.29-80.61\\
0.01& 49.64& 33.57-65.71 & 154.71 &104.77-204.65\\
\\
[0.5ex]
\end{tabular} 
\caption{Values of the product $\langle\sigma\upsilon\rangle$, in cm$^3$\,s$^{-1}$, which give rise to 
positron fractions which best-fit the HEAT observations, for the four MSSM models considered, for 
$M_{\rm cut-off}=10^6$ and various values of $c$ and $f$. One sigma error ranges are also 
displayed.} 
\label{table2} 
\end{center} 
\end{table}   

\section{Assessment}
The results displayed in figures\,\,\ref{figure2}-\ref{figure5} indicate that neutralino annihilations  
within local dark matter substructure can give rise to positron fractions which correspond well with observations,  
with $\chi^2$ between 2.9 and 5.4, for values of $\langle\sigma\upsilon\rangle$ ranging from $10^{-29}$cm$^3$s$^{-1}$ for 
50\,GeV neutralinos to $2\times10^{-27}$cm$^3$s$^{-1}$ for 600\,GeV neutralinos.   

Such values of $\langle\sigma\upsilon\rangle$ are at least an order of magnitude smaller than the canonical estimate of 
$3\times10^{-26}$\,cm$^3$s$^{-1}$, determined from relic density calulations (\cite{bertone05}).  
However, many conventional MSSM models exist in which this approximation is grossly violated and are easily able to 
accomodate the results obtained, where the number of such models rapidly increases for lighter neutralinos, 
especially near 50\,GeV (\cite{hooperb}).

As expected, we observe that the best-fit values of $\langle\sigma\upsilon\rangle$ increase as the dark matter clumps 
become increasingly stripped, and decrease as we increase their concentration, where the latter makes sense since a 
larger concentration implies  that an increasing proportion of the clump mass is contained within a smaller radius.  
However, we observe that even with up to 99\% stripping, and varying the clump concentration over the entire range 
observed by Diemand et al., the best-fit values of  $\langle\sigma\upsilon\rangle$ do not change by more than an order 
of magnitude.  

We note that, of course, in reality the density of the clumps cannot diverge at their centres, as  
an NFW profile suggests, but are likely to possess uniformly dense cores containing of order 1\% of  
their total mass (\cite{zhao05}).  The stripping and eventual disintigration of these cores is a topic of 
much debate and is currently being  investigated (\cite{green}). We therefore do not display  results for 
$f<0.01$, since it is unknown as to whether these cores can survive such severe levels of  disruption.  

However, in table\,\ref{table2} we display results similar to those in table\,\ref{table1},  
now for $M_{\rm cut-off}=10^6\,M_{\odot}$, corresponding  to the typical mass resolution of simulations of 
the galactic halo, in order to simulate the destruction of lighter clumps. We observe that the best-fit values 
of $\langle\sigma\upsilon\rangle$ increase by no more than an order of magnitude compared to their corresponding 
values in table \ref{table1}, but are more consistent with the canonical value and consequently consistent with a 
larger number of MSSM models.  

We should also note that these results may be slightly modified when acknowledging flux contributions from 
tidal streams resulting from material stripped from clumps (\cite{zhao05}). However, owing to the significantly lower 
densities of these streams, compared to the clumps from which they originate, it is very likely that such effects 
will not significantly alter our results.

\section{Conclusions}
For each of the four benchmark supersymmetric models considered, we solved the diffusion-loss equation for the local 
positron flux resulting from neutralino annihilations within local dark matter substructures. We utilised the results 
from the unprecedentely high resolution simulation conducted by Diemand et al., who determined that the lightest
clumps to form are of order an Earth mass and as large as $10^{-2}\,$pc. The mass distribution for clumps
resulting from the simulation, which was terminated at $z=26$, has been shown to be similar to be that expected today 
for large mass clumps. We assigned each clump an NFW density profile, with an associated universal concentration parameter 
consistent with the numerical results. We investigated the effects on the flux when invoking different degrees of 
(universal) mass loss suffereed by each clump, in order to simulate the effects of tidal-stripping from stellar encounters.

Our results indicate that neutralino annihilations  
within local dark matter substructure can give rise to positron fractions which correspond well with the observations by 
HEAT, for values of $\langle\sigma\upsilon\rangle$ ranging from $10^{-29}$cm$^3$s$^{-1}$ 
for 50\,GeV neutralinos to $2\times10^{-27}$cm$^3$s$^{-1}$ for 600\,GeV neutralinos.  
Despite that such $\langle\sigma\upsilon\rangle$ are at least an order of magnitude smaller than the canonical value of 
$3\times10^{-26}$\,cm$^3$s$^{-1}$, determined from relic density calulations we indicate that many conventional MSSM models 
exist in which this approximation is grossly violated and are easily able to accomodate our results obtained, where the 
number of such models rapidly increases for lighter neutralinos, especially near 50\,GeV.

We observed that even with up to 99\% stripping, and varying the clump concentration over the entire range 
observed by Diemand et al., the best-fit values of $\langle\sigma\upsilon\rangle$ do not change by more than an order 
of magnitude. 
We also observed that even if we change the lower mass cut-off in our mass distribution from an Earth mass to $10^{6}$ solar
masses, the best-fit values of $\langle\sigma\upsilon\rangle$ increase by no more than an order of magnitude, but are consequently
more consistent with a larger region of MSSM parameter space.  

\section*{Appendix:\:\:\:Derivation of the steady-state positron flux}
Here we derive in detail the steady-state solution to the diffusion-loss equation (\ref{1}), given by

\begin{eqnarray} \frac{\partial}{\partial t} \frac{\partial n_{e^+}}{\partial E_{e^+}} & = &  
\nabla \cdot \left[K(E_{e^+},{\bf r})\nabla\frac{\partial n_{e^+}}{\partial E_{e^+}}\right] \nonumber\\
& & +\frac{\partial}{\partial E_{e^+}} \left[b(E_{e^+},{\bf r})\frac{\partial n_{e^+}}{\partial E_{e^+}}\right] + 
Q(E_{e^+},{\bf r}).
\label{A1} 
\end{eqnarray}   

\noindent Our treatment is very similar to that described in \cite{edsjo98}, except that several unnecessary assumptions
made in that derivation are omitted here, making our solution slightly more general.

We firstly re-arrange (\ref{A1}) in terms of the dimensionless parameter $u=1/\epsilon=1\,{\rm GeV}/E_{e^+}$,
and re-write the diffusion coefficient, given by (\ref{2}), as $K=K_0h(u)$, which gives

\begin{equation}
\frac{1}{h(u)}\frac{\partial}{\partial u}\frac{\partial n_{e^+}}{\partial u}=K_0\tau_E\nabla^2\frac{\partial n_{e^+}}{\partial u}-\tau_E\left[u^2h(u)\right]^{-1}Q(\epsilon(u), {\bf r}), 
\label{A3}
\end{equation}

\noindent where we have used equation (\ref{3}) for the energy-loss rate, $b$. We now re-express (\ref{A3}) in terms of the
variable $v$, where $h(u)=dv/du$. Provided that $h(u)$ is differentiable, we obtain

\begin{equation}
\frac{\partial}{\partial v}\frac{\partial n_{e^+}}{\partial u}=K_0\tau_E\nabla^2\frac{\partial n_{e^+}}{\partial u}-\tau_E\left[u^2h(u)\right]^{-1}Q(\epsilon(v), {\bf r}).
\label{A4}
\end{equation} 

\noindent Now making the substitutions

\begin{equation}
w(v)=-\frac{d\epsilon}{dv}=\left[u^2h(u)\right]^{-1},
\label{A6}
\end{equation}

\noindent and

\begin{equation}
F(v,{\bf r})=\frac{\partial n_{e^+}(v,{\bf r})}{\partial u(v)},
\label{A7}
\end{equation}

\noindent equation (\ref{A4}) becomes

\begin{equation}
\frac{\partial }{\partial v}F(v,{\bf r})=K_0\tau_E\nabla^2F(v,{\bf r})-\tau_Ew(v)Q(\epsilon(v), {\bf r}).
\label{A8}
\end{equation} 

As stated in \cite{edsjo98}, equation (\ref{A8}) is analogous to an inhomogeneous heat equation. The spatial variables
are exactly analogous, while $v$ is analogous to "time". Since $v=\int h(u)du=3^{\alpha}u+(1-\alpha)^{-1}u^{1-\alpha}$,
we find that $v$ is a monotonically increasing function for $\alpha <1$, and therefore is appropriately analogised to time.

Considering the above, we can write the solution of the steady-state positron number density as an integral over the
effective source term, $-\tau_Ew(v)Q(\epsilon (v), {\bf r})$, of equation (\ref{A8}), multiplied by a Green's function.

We firstly solve for the free Green's function $G_{{\rm free}}$, which satisfies

\begin{eqnarray}
\frac{\partial}{\partial v}G_{{\rm free}}(v-v', {\bf r}-{\bf r'})&-&K_0\tau_E\nabla^2G_{{\rm free}}(v-v', {\bf r}-{\bf r'})\nonumber\\
&=&\delta(v-v')\delta({\bf r}-{\bf r'}),
\label{A9}
\end{eqnarray}

\noindent which is identical to equation (\ref{A8}) with an effective source term $\delta(v-v')\delta({\bf r}-{\bf r'})$,
which is proportional to the source term for a monoenergetic point source, with $v=v'$, located at ${\bf r}={\bf r'}$. Taking the
(spatial) Fourier transform of (\ref{A9}), we obtain

\begin{eqnarray}
\frac{\partial}{\partial v}\tilde{G}_{{\rm free}}(v-v', {\bf k})&+&K_0\tau_E\nabla^2k^2\tilde{G}_{{\rm free}}(v-v', {\bf k})\nonumber\\
&=&(2\pi)^{-3/2}\delta(v-v'){\rm exp}(-i{\bf k}\cdot{\bf r'}),\nonumber\\
\label{A10}
\end{eqnarray}

\noindent where $\tilde{G}_{{\rm free}}$ is the Fourier transform of $G_{{\rm free}}$. Multiplying each side of 
(\ref{A10}) by the integrating factor $I={\rm exp}(K_0\tau_Ek^2v)$ and integrating over $v$, we obtain

\begin{eqnarray}
\tilde{G}(v-v',{\bf k})&=&(2\pi)^{-3/2}{\rm exp}(-K_0\tau_Ek^2(v-v'))\nonumber\\
&&\times{\rm exp}(-i{\bf k}\cdot{\bf r})\theta(v-v'),
\label{A11}
\end{eqnarray}

\noindent where we have invoked the condition $G_{\rm free}=0$ (and therefore $\tilde{G}_{\rm free}=0$) for $v=0$ 
(i.e. at infinite energy). Finally, we take the inverse Fourier transform of (\ref{A11}), to obtain

\begin{equation}
G_{{\rm free}}(v-v', {\bf r}-{\bf r'})=(2\pi)^{-3/2}\int_{\bf k}\tilde{G}(v-v', {\bf k}){\rm exp}(i{\bf k}\cdot{\bf r})d^3{\bf k},
\label{A12}
\end{equation}

\noindent which evaluates to

\begin{eqnarray}
G_{{\rm free}}(v-v', {\bf r}-{\bf r'})&=&\left[D(v,v')\right]^{-3}{\rm exp}\left(\frac{({\bf r}-{\bf r'})^2}{\left[D(v,v')\right]^2}\right),\nonumber\\
\label{A13}
\end{eqnarray}

\noindent where $D(v, v')=\left[4\pi K_0\tau_E(v-v')\right]^{1/2}$. 

As described in $\S\,3$, our chosen diffusion zone is an infinite cylindrical slab of thickness $2L$,
where positrons located outside of this region simply free-stream. Therefore the steady-state positron
distribution must therefore vanish at $r=\infty$ and $|z|>L$. Ignoring contributions form sources outside of the 
diffusion zone, then from the form of the diffusion equation, the positron distribution will be zero for 
$|z|>L$ if it is zero at $z=\pm L$. Therefore in order to obtain our desired positron distribution within the diffusion 
zone, we can use a Green's function which simply vanishes at the boundaries of our diffusion zone. To do this we adopt the 
solution proposed by \cite{edsjo98}, which utilises a set of image charges, 

\begin{equation}
x'_n=x',\:\: y'_n=y',\:\: z'_n=2Ln+(-1)^nz',
\label{A14}
\end{equation}

\noindent to find the required Green's function,

\begin{equation}
G_{2L}(v-v', {\bf r}-{\bf r'})=\sum_{n=-\infty}^{\infty}(-1)^nG_{{\rm free}}(v-v', {\bf r}-{\bf r'}_n).
\label{A15}
\end{equation}

\noindent As described above, the steady-state positron distribution can then be expressed as the following integral

\begin{eqnarray}
\frac{\partial n}{\partial\epsilon}&=&\tau_E\epsilon^{-2}\int_0^{v(\epsilon)}dv'w(v')\nonumber\\
&&\times\int d^3{\bf r'}G_{2L}\left(v(\epsilon)-v', {\bf r}-{\bf r'}\right)Q(\epsilon(v'), {\bf r'}).
\label{A16}
\end{eqnarray}

\noindent Then using equation (\ref{A6}), we can conveniently re-express (\ref{A16}) as

\begin{eqnarray}
\frac{\partial n}{\partial\epsilon}&=&\tau_E\epsilon^{-2}\int_{\epsilon}^{\infty}d\epsilon'\nonumber\\
&&\times\int d^3{\bf r'}G_{2L}\left(v(\epsilon)-v'(\epsilon'), {\bf r}-{\bf r'}\right)Q(\epsilon, {\bf r'}).
\label{A17}
\end{eqnarray}

Now we must construct an appropriate source function, $Q$. To do this we only consider contributions from dark matter 
substructure, ignoring the contribution from the smooth component of the halo, which we calculate to be approximately $10^{-4}$ times the 
substructure
component. As described in $\S\,4$, we assume each clump to have an NFW profile, $\rho_{\rm NFW}(c,r_{200}(M))$, with
concentration, $c$, and maximum radius, $r_{200}(c,M)$, determined by c and the clump mass, $M$. The rate of production of positrons
of energy $\epsilon\,$GeV, is then

\begin{equation}
\frac{\langle\sigma\upsilon\rangle}{m_{\chi}^2}\frac{d\phi}{d\epsilon}\int_0^{r_{200}(c,M)}\rho^2(r')4\pi r'^2dr'
=\frac{\langle\sigma\upsilon\rangle}{m_{\chi}^2}\frac{d\phi}{d\epsilon}f^2_{{\rm NFW}}(c,M)
\label{A18}
\end{equation}

\noindent where $\langle\sigma\upsilon\rangle$ is the thermally-averaged, low-velocity annihilation cross-section
multiplied by relative speed and $d\phi/d\epsilon$ is the number of positrons produced per annihilation per energy
interval.

We must now multiply equation (\ref{A18}) by the correctly normalised density of clumps. As described in $\S\,4$ we
utilise the clump distribution, $dn_D(M)/d{\rm log}M$, determined by Diemand et al., normalised to a local clump density of $500\,$pc$^{-3}$, 
and like Diemand, we assume that this normalisation is proportional to the underlying halo density profile, 
$\rho_{\rm halo}$. For simplicity, we choose the halo profile to be a 'cylindrically' symmetric NFW profile

\begin{equation}
\rho_{{\rm halo}}(r)=\frac{\rho_0}{(r/R)(1+(r/R))^2}
\label{A19}
\end{equation}

\noindent where $r$ is the cylindrical radial coordinate, $R\approx 20\,$kpc and $\rho_0$ is a constant. The correctly 
normalised, spatially-dependent mass distribution of clumps is then

\begin{eqnarray}
\frac{dn(r,M)}{d{\rm log}M}&=&\frac{\rho_{{\rm halo}}(r)}{\rho_{{\rm halo}}(r=R_{\odot})}\,\frac{dn_D(M)}{d{\rm log}M}\nonumber\\
&\simeq&\frac{0.86}{(r/R)(1+(r/R))^2}\,\frac{dn_D(M)}{d{\rm log}M},
\label{A20}
\end{eqnarray}

\noindent where $R_{\odot}\approx 8.5\,$kpc, is the approximate distance between the Earth and the galactic centre. 

Considering the above, the source function is given by

\begin{equation}
Q(\epsilon, r)=\frac{\langle\sigma\upsilon\rangle}{m_{\chi}^2}\frac{d\phi}{d\epsilon}\int^{M_{\rm max}}_{M_{\rm min}}f^2_{{\rm NFW}}(c,M)\frac{dn(r,M)}{d{\rm log}M}d{\rm log}M,
\label{A21}
\end{equation}  

\noindent where $M_{\rm min}\sim 10^{-6}M_{\odot}$ and $M_{\rm max}\sim 10^{10}M_{\odot}$, as discussed in $\S\,4$.

Substituting expressions (\ref{A15}), (\ref{A21}) and (\ref{A20}) into equation (\ref{A17}), evaluated at $z=0$ and $r=R_{\odot}$,
we obtain the required (local) positron number density

\begin{eqnarray}
\left(\frac{\partial n_{e^+}}{\partial \epsilon}\right)_{{\rm local}}&=&\frac{\langle\sigma\upsilon\rangle\tau_E}{(m_{\chi}\epsilon)^2}
\int\limits^{M_{\rm max}}_{M_{\rm min}}f^2_{{\rm NFW}}(c,M)\frac{dn_D(M)}{d{\rm log}M}d{\rm log}M\nonumber\\
&&\times\int\limits^{\infty}_{\epsilon}d\epsilon'\left\{D[v(\epsilon), v'(\epsilon')]\right\}^{-3}\frac{d\phi}{d\epsilon'}\sum^{\infty}_{n=-\infty}(-1)^n\nonumber\\
&&\times\int\limits^L_{-L}dz'{\rm exp}\left[-\left(\frac{(-1)^nz'+2Ln}{D\left[v(\epsilon), v'(\epsilon')\right]}\right)^2\right]\nonumber\\
&&\times\int\limits_0^{\infty}dr'r'\frac{0.86}{(r'/R)(1+(r'/R))^2}\nonumber\\
&&\times{\rm exp}\left(-\frac{R_{\odot}^2+r'^2}{\left\{D\left[v(\epsilon), v'(\epsilon')\right]\right\}^2}\right)\nonumber\\
&&\times\int\limits_0^{2\pi}d\theta'{\rm exp}\left(\frac{2rr'{\rm cos}\theta'}{\left\{D\left[v(\epsilon), v'(\epsilon')\right]\right\}^2}\right).
\label{A22}
\end{eqnarray}

\noindent The $\theta'$ integral evaluates to $2\pi I_0$, where $I_0$ is a modified bessel function of the first kind and
the $z'$ integral evaluates to the following sum of error functions,

\begin{eqnarray}
H(L,D)&=&\pi^{1/2}D\bigg({\rm erf}\left(\frac{L}{D}\right)+\sum^{\infty}_{m=1}\bigg\{{\rm erf}\left[\frac{(4m-3)L}{D}\right]\nonumber\\
&&+{\rm erf}\left[\frac{(4m+1)L}{D}\right]-2{\rm erf}\left[\frac{(4m-1)L}{D}\right]\bigg\}\bigg)\nonumber\\.
\label{A23}
\end{eqnarray}

\noindent We are then left with integrals over $\epsilon'$ and $r'$,

\begin{eqnarray}
\left(\frac{\partial n_{e^+}}{\partial \epsilon}\right)_{{\rm local}}&=&\frac{2R\langle\sigma\upsilon\rangle\tau_E}{(m_{\chi}\epsilon)^2}
\int\limits^{M_{\rm max}}_{M_{\rm min}}f^2_{{\rm NFW}}(c,M)\frac{dn_D(M)}{d{\rm log}M}d{\rm log}M\nonumber\\
&&\times\int\limits^{\infty}_{\epsilon}d\epsilon'\left[D(v, v')\right]^{-2}\frac{d\phi}{d\epsilon'}H(L,D(v,v'))\nonumber\\
&&\times\int\limits_0^{\infty}dr'\frac{0.86}{(1+(r'/R))^2}\nonumber\\
&&\times{\rm exp}\left(-\frac{R_{\odot}^2+r'^2}{\left[D\left(v, v'\right)\right]^2}\right)\nonumber\\
&&\times I_0\left(\frac{2R_{\odot}r'}{\left[D\left(v, v'\right)\right]^2}\right),
\label{A24}
\end{eqnarray}

\noindent which we evaluate numerically.

Finally, using equation (\ref{A24}), we obtain the local positron flux resulting from neutralino annihilations to be

\begin{equation}
\left(\frac{d\Phi_{e^+}}{d\epsilon}\right)_{{\rm local}}=\frac{\beta c}{4\pi}\left(\frac{dn_{e^+}}{d\epsilon}\right)_{{\rm local}},
\label{A25}
\end{equation}

\noindent where $\beta c$ is the speed of a positron of energy $\epsilon$.

\section*{Acknowledgements}
We would like to thank S. Nussinov and D. Hooper for various discussions with us on  this topic. 


\label{lastpage}
\end{document}